# Gender-Related Systematics in the NRAO and ALMA Proposal Review Processes

Carol J. Lonsdale, Frederic R. Schwab and Gareth Hunt

National Radio Astronomy Observatory, 520 Edgemont Road, Charlottesville, VA 22903

**Abstract**

A study has been made of the evidence for gender-related systematics in the proposal review processes for the four facilities operated by NRAO: the Jansky Very Large Array (JVLA; hereafter VLA), the Very Long Baseline Array (VLBA), the Green Bank Telescope (GBT) and the Atacama Large Millimeter/submillimeter Array (ALMA) in Chile which is operated by NRAO/AUI in partnership with the European Southern Observatory (ESO) and the National Astronomical Observatories of Japan (NAOJ), in cooperation with the Republic of Chile. A significant gender-related effect is found in the proposal rankings in favor of men over women in the ALMA Proposal Review Processes (PRP) for ALMA Cycles 2-4, with reliability of 99.998% that the underlying rank distributions for male and female PIs are not the same. The effect is largest and most significant for ALMA Cycle 3. A similar overall result is found for the other three NRAO telescopes over proposal Semesters 2012A-2017A, but with lower reliability level overall (98.3%), and with some reversals across semesters in the trend for better performance in the rankings for male PIs. The results align with similar studies recently completed for the HST (Reid 2014) and the ESO proposal review processes (Patat 2016). No correlations are found between the gender-related proposal ranking trends and the gender fractions on review panels. The HST and ESO proposal reviews have come to different conclusions from each other on the role of seniority on the gender-related proposal outcomes at those observatories. The currently available data for the ALMA and NRAO user base do not allow us to investigate the important question of the dependence on the gender-related trends of the seniority of the principal investigators.

## I. Introduction

Reid (2014) reports a study of gender-based systematic trends in the HST proposal review process for Cycles 11 through 21. Key results from this study are that there appears to be a systematic trend that male PIs are more likely to succeed in achieving a successful HST proposal than female PIs. The effect is not necessarily significant in a single cycle but the pattern is consistent. No evidence was found for a dependence of the bias on geographic origin of the proposal or on gender distribution of the review panels, however the resulting bias in the success rates does appear to ameliorate for more recent female graduate proposers and possibly also with an increase in the proportion of more junior scientists on the review panels, suggesting a generational effect disfavoring the more senior women. Interestingly there is no strong dependence on the panel science topic, however Reid found that the same panel could produce different gender-based success rates within different science categories. A clue is also present that the gender-related trends arise in the face-to-face panel meetings because there was less evidence for the systematic trends amongst triaged proposals.

Patat (2016) has undertaken a similar study of the time allocation process at the European Southern Observatory (ESO), covering a period of 8 years and involving about 3000 principal investigators. He finds that female PIs have a significantly lower chance of being awarded a top rank compared to male PIs and suggests that the principal explanation for the difference may be due to the average higher seniority of the male PIs, if the assumption is made that more senior scientists of both genders write better proposals and thus succeed more often at getting ESO telescope time. He also finds evidence for an additional "gender-dependent behavior" in the proposal review patterns for both male and female reviewers, with evidence for a somewhat more pronounced effect amongst male reviewers.

The NRAO is deeply committed to principles of equity and inclusion in all aspects of its operation as an observatory and as a community, therefore a study on the possibility of gender-related systematics in the review processes of proposals to the four telescopes operated by NRAO and its partners has been undertaken. The results of the study are reported here.

The interpretation of observed trends that are apparently gender-related is complex. In addition to actual gender bias (conscious or unconscious discrimination in favor of one gender over another) the field of astrophysics is subject to other kinds of biases, as is true in society as a whole. In particular, when considering gender-based systematics, it is important to also understand the concept of prestige bias. Prestige bias arises when a figure with a more prestigious reputation is accorded more consideration than a less well-known figure. Indeed some level of prestige bias is built into the proposal review process of many telescopes since review panels are often directed to take into account the record of the team and ability of the team to accomplish the project. This has the effect of disfavoring proposals by younger and less well-known researchers. Gender and prestige bias are obviously interconnected in any field in which one gender is in the minority, especially at the higher levels of recognition, therefore prestige bias may compound, or masquerade as, gender bias. The same complex situation applies to minorities of all kinds because a minority most likely has fewer individuals who are accorded the higher levels of esteem. Such a situation could arise in astronomy, for example, when a relatively new scientific community enters an established telescope user base.

It is this effect, prestige bias, that Patat (2016) may be finding evidence for in his ESO study, while Reid (2014) may be seeing a more complex scenario in the HST proposal result patterns, which could in fact run counter to the effects of prestige bias. In this study we use the term "gender-related" to encompass any and all conscious or unconscious biases that may be contributing to an apparent imbalance in the success rate of male and female proposers to the NRAO telescopes. A more detailed investigation into the complex factors that might be at play to cause the observed effects will be addressed in a future publication.

## II. The ALMA and NRAO Proposal Review Processes

NRAO proposal review for the VLA, VLBA and GBT occurs twice yearly and is done by eight Science Review Panels (SRPs), covering eight distinct science categories and each comprising six members. Before meeting face-to-face each (non-conflicted) member of a panel submits a score for every proposal assigned to that panel on a scale of 0.1 (best) to 9.9 (worst). The chair submits a score only if there are one or more conflicts. The scores submitted by each reviewer are normalized, and then the scores for each proposal are averaged to produce a mean and a standard deviation. The SRP then meets by telecon, and members discuss proposals and can alter the scores. After the conclusion of the SRP meetings the scores from each panel are then "linearized", and then merged together to obtain a final rank-order list that is submitted to the Time Allocation Committee (TAC), whose members are the chairs of the SRPs. The TAC is responsible for setting the observation priorities of the NRAO proposals, taking into account weather, RA, observation frequency, etc.

The ALMA Proposal Review Process proceeds in a similar fashion. The ongoing ALMA observing cycles now proceed on a yearly basis, however the first three cycles, Cycles 0, 1 and 2, had varying lengths due to the Early Science nature of those cycles, and the review procedures for Cycles 0 and 1 differed somewhat from those of the later cycles. Here we focus on the results from Cycles 2, 3 and 4, which had uniform procedures. The proposals are divided into 5 science categories each having 1 to 3 separate review panels called ALMA Review Panels (ARPs) consisting of 7 to 8 members each. Each panel produces a rank-ordered listing of proposals that is then linearized and combined into one unique rank-ordered list by the Proposal Handling Team (PHT), which is then considered by the ALMA Proposal Review Committee (APRC), consisting of the chairs of all the ARPs. Before the ARP panels meet in person, each proposal is assigned to a subset of the ARP members, who provide a score from 1 to 10 (10 being worst). The scores are normalized and means computed, and a single-ranked list created. A triage line is drawn and only proposals above the line proceed to the panels.

The role of the APRC is to review the results for the ARPs and to generate a final, single, ranked list. The process differs at this stage from that for NRAO in that the APRC does not take into account weather or sky availability but only science value. Once the final rank-ordered list is made, the Joint ALMA Observatory (JAO) then has the responsibility of generating the grades, which take into account partner balance, sky availability, predicted weather patterns including day vs. night observing (affecting high frequency observability), and configuration availability.

Our starting point for the analysis is the final rank-ordered lists before the grades are assigned, to avoid the non-gender-based systematics inherent in the grades. This differs from the HST analysis of Reid that used proposal success as a statistic.

## III. Data and Methods

The data primarily used for this analysis include proposals submitted for ALMA Cycles 2-4 and NRAO proposal semesters 2012A-2017A. The total number of proposals involved in the analyses is 4523 ALMA proposals from 1762 PIs, and 3862 proposals to VLA, GBT and VLBA telescopes from 1382 PIs. We also review results for 940 proposals from 361 PIs for ALMA Cycles 0 and 1, but do not combine these data with the Cycle 2-4 data due to differences in the review procedures. Specifically, only the top 18% of the Cycle 0 proposals retain a rank in the records, and for Cycle 1 only the top 70% of the proposals retain a rank.

As neither NRAO nor ALMA proposers are required to provide gender information, the appropriate association (Female or Male) was made for each PI using information researched on the Web for cases in which the PI did not provide this information. Gender could not be assigned unambiguously in this way for 19 ALMA and for 3 NRAO PIs, and these proposals were consequently not included in the analysis. Unlike the ESO study, the career status of the ALMA and NRAO PIs is not currently available, therefore no analysis related to the seniority of the PIs is possible at this time.

Given the merged normalized ranks for each telescope we compare the distributions of the proposals with female and male PIs (F and M). We have used the Kolmorgorov-Smirnov (KS) Test and the more reliable Anderson-Darling (AD) Test (Babu & Feigelson 2006) to evaluate the probability that the observed rank distributions for the male and female PIs are drawn from the same parent sample. The AD results are listed in column 2 of Table 1 for ALMA Cycles 2-4 and of Table 3 for VLA, GBT and VLBA. If the null hypothesis that the male and female ranking distributions are drawn from the same parent distribution is rejected at the 80% or higher confidence level, the table cell has been colored in purple, with the shade deepening as the confidence level goes from 80-90% to 90-95% to 95-99% to >99%.

The results for ALMA Cycles 0 and 1 are illustrated separately in Table 2, because only the top 18% of the Cycle 0 proposals, and the top 70% of the Cycle 1 proposals, have a rank.

We also compare the value of the rank for male and female PIs at several percentiles of the overall distributions in Tables 1, 2 and 3. We present the ranks at the $25^{th}$, $50^{th}$ (the median) and $75^{th}$ percentile. For ALMA Cycles 1-4 the $25^{th}$ percentile is of most relevance to the success of ALMA proposals due to the high over-subscription rates, while for ALMA Cycle 0 all three quartiles are of interest since the ranks only represent the most successful 18% of the proposals. For NRAO the $25^{th}$ and $50^{th}$ percentiles are relevant for proposal success.

In order to also generate confidence levels on the F and M ranks at the $25^{th}$, $50^{th}$ and $75^{th}$ percentiles (also referred to, more simply, as the "quartile values"), the quartile ranks and their reliability were computed by a resampling method. We use the "smoothed bootstrap" (Efron 1992) method within Mathematica, constructing a kernel-density estimator to fit each quartile data subset then generating 10,000 random samples (replicates) from these model distributions. The default Gaussian kernel was used, whose characteristic width is chosen automatically, depending on the nature of the data. The results of this analysis are illustrated in Figures 1 and 2, and a more detailed report for the combined data for ALMA (Cycles 2-4 combined) and for the NRAO telescopes (all 3 telescopes for all 2012-2016 semesters combined) is provided in the Appendix.

The ALMA observatory is supported by three executives, North America (NA), East Asia (EA) and Europe (EU), and is hosted by Chile (CL). NA and EU proposers are awarded 33.75% of ALMA time each, EA is awarded 22.5% and Chile has 10%. Proposals from the small number of PIs from none of the regions are included with the NA share. Due to the renormalization required to achieve these percentages, the final grade outcomes (A, B, C and U) of the proposal review process are region-dependent. For example the A grades for one region might be awarded to proposals with pre-linearized ranks between 1 and 25 but between ranks 1 and 50 for a smaller region. However the final linearized ranked list produced by the ALMA APRC does not consider region, and therefore we can compare directly the distributions for the 4 regions. These results are summarized in Table 1.

## IV. Results

### a. ALMA

#### i. Combined Data

For ALMA Cycles 2-4 and all regions combined, the overall rank distributions for male and female PIs are shown to not be drawn from the same parent population with a high level of confidence by the Anderson-Darling (AD) tests. This is highlighted in Table 1 (column 2): the combined Cycle 2+3+4 results show that the rank distributions for male and female PIs are different with >99.99% confidence. Figure A1 in the appendix illustrates the sense of this difference: the histogram of male PI ranks falls steadily towards the poorer ranks while the histogram for female PIs rises.

The results of the quartile analysis for the cycle- and region-combined data confirm the clear trend for the ranks of the men to be systematically better than those of the women across the full range of the ranks, shown clearly in Figure 1, top panel. All three quartiles show highly significant confidence, reaching over 99%, that the ranks are better for male PIs.

#### ii. ALMA Data Breakdown by Cycle

Considering Cycles 2, 3 and 4, the data when broken down by cycle show that the overall rank distributions trend towards favoring male PIs, with the strongest signal from the AD test for Cycle 3: >99.99% confidence. The AD test results are less reliable for Cycle 2, with ~90% confidence, and the trend may ameliorate in Cycle 4, where the significance drops to a confidence level of only ~70%. This latter result is particularly interesting in light of the fact that the ALMA Cycle 4 review panel members were alerted to the gender-related trends present in the Cycle 2 and 3 data and accordingly were asked to keep the goal of gender neutrality centrally in mind during the Cycle 4 review deliberations.

Still considering Cycles 2, 3 and 4, the quartile analysis results for the separate cycles, illustrated in Figure 1, confirm that the trend seen in the combined data for the ranks of the men to be systematically better than those of the women, across the full range of the ranks, is also present for each cycle. In the individual cycles, the most significant male-female PI rank separation in the quartile data is for Cycle 3, for which all three quartiles show the male PI ranks exceeding those of female PIs at well above the 95% confidence level. For Cycles 2 and 4 the separations are less significant, with levels reaching only ~68% significance favoring male PIs. Even though the significance of each the individual quartile measurements is lower for Cycles 2 and 4, the fact that most quartiles for both of these cycles show male dominance reinforces the conclusion that a trend related to gender is likely present in every cycle. This is best seen from the trends in blue vs. tan shading of the cells in Table 1.

Turning to Cycles 0 and 1 we have limited information because only the top 18% and 70% of the proposals, respectively, have ranking information. Moreover, the Cycle 0 data are affected by the considerations that the APRC used to assign final grades, since only proposals that were ultimately assigned High Priority or Filler status have ranks for Cycle 0. For Cycle 1 the grade assignment process is not a concern since it is only the 30% of proposals that were triaged that do not have ranks. In spite of the limitations we present the results for the sake of completeness and of transparency.

From Table 2 there is some evidence that the top-ranked 18% of proposals may tend to favor female vs. male PIs in Cycle 0 (top panel) though the significance levels are not high. Considering Cycle 1 (bottom panel of Table 2) there is not a strong signal indicating male dominance overall among the 70% of proposals that were not triaged.

#### iii. ALMA Data Breakdown by Region

We have examined the trends by region for the three Cycles with consistent data sets: Cycles 2, 3 and 4. The same trends seen in the combined data are also generally seen for each individual region (Table 1). In principle, this would be expected since the same SPRs and APRCs review the proposals from all regions. The significance is naturally expected to be reduced when the data are divided in this way however the trends remain strong in many cases. The highest confidence level results are for NA in Cycle 3, reaching almost 95%, closely followed by EU in Cycle 3. An exception is EA where the significance of any trend is small, a result that is not entirely attributable to lower number statistics and may be a real effect. An analysis of the possible causes of regionally-related trends is beyond the scope of this article, however the effect of culturally-related issues would be of interest in a future study.

### b. NRAO

For the NRAO telescopes we have derived results for all instruments combined and also for each separately, shown in Table 3. We have also derived the data by year beginning with 2012, and by semester, as well as for all years combined.

#### i. NRAO Combined Data

When all NRAO telescopes and all years (2012-2016) are combined, the significance that the parent population is not the same for the M and F PI ranks is about 98% from the AD test. There is some evidence that the rank distribution for men is wider and flatter than for women, for whom it is peaked in the middle ranks (histograms in Figure A2). This is in contrast to ALMA where we see the number of women increasing steadily towards the poorer ranks and the number of men increasing steadily towards the better ranks (Figure A1). In spite of this difference in overall ranking distribution shapes, male PIs have somewhat better first and second quartile ranks than female PIs (see Table 3) and therefore somewhat better overall proposal outcomes. The trend is both smaller and less significant than for ALMA.

#### ii. NRAO Data Breakdown by Year and by Semester

The results of the AD tests for the separate years (Table 3) indicate that differences between the overall male and female PI rank distributions for NRAO are not as significant as they are for ALMA. There is only one year for which the AD test confidence level is higher than 95%, 2014, and it is in the sense of men outperforming women. For 2013 and 2016 the AD test confidence levels are above 80%. For 2016 the evidence is for male dominance, however for 2013 the trend is for female dominance.

Looking at the quartile results, and considering only the 25th and 50th quartiles as relevant to proposal success, there are only two cases for which the difference between male and female PI quartile ranks is significant above the 68% confidence level. These are the 2014-25th and 2016-25th percentiles, where the male PIs outperform female PIs in both cases. In both years the 50th percentiles show the same male dominance with lesser significance. For 2013, the other year noted above with a possibly significant AD test result (favoring female PI dominance), the quartile results are consistent with female dominance.

The data have also been broken down further into 6-month semesters, which naturally causes the uncertainties of the analysis to increase. We also include the results for the most recent semester, 2017A. There are two semesters with a high significance level for a difference between the male and female rank distributions, 2014B (95%) and 2016B (98%), despite the lower number statistics. 2017A also shows an indication of a gender-related signal at 87% confidence. In all three cases the male PIs have better quartile ranks than the female PIs. For 2014B the result confirms that for the full 2014 year, while for 2016B the result has higher confidence than the full 2016 year.

### iii. NRAO Data Trends with SRP Gender

Due to the fact that some possible reversals are seen in gender dominance for the NRAO telescopes, the semester data trends can be compared to the changing fraction of women members of the NRAO SRPs, which ranges from 18.7% (Cycle 15A) to 27% (Cycles 2013B and 2014A). There is no clear correlation between this fraction and any gender-related trends.

We also looked at the trends in SRP membership for the two semesters which experienced the largest shifts towards significant outperformance by male PIs, 2014B and 2016B, since it is plausible that a shift in proposal ranking trends might be more likely when there is a larger rotation in SRP membership from one semester to the next. This was done anonymously, replacing SRP member names with numbers by one of us, and the analysis done by another. Specifically, we show in Table 4 the continuation pattern for SRP members from one NRAO semester to the next. The data show no evidence to support such a scenario.

### iv. VLA, VLBA and GBT Considered Separately

The trends found for the different NRAO telescopes individually have lower significance due to low number statistics. From the Anderson-Darling test results we find that the confidence level on a real difference between the male and female rankings is ~90% for both VLA and GBT when all years are combined. In both cases the sense is that of men outperforming women, as confirmed by the quartile results. The VLBA results are not statistically significant.

**TABLE 1: ALMA Cycles 2-4: AD Tests for Rank Distributions and Ranks at the Three Quartiles.**

| ALMA | Anderson-Darling Test p-value[1] | NUMBER OF PIS | | % F PIs | 25th percentile | | 50th percentile | | 75th percentile | |
|---|---|---|---|---|---|---|---|---|---|---|
| | | F | M | | F-M[2] | CL[3] | F-M | CL | F-M | CL |
| | By Cycle & Summed, all Regions Combined | | | | | | | | | |
| Cycle 2 | 0.099 | 412 | 915 | *31.1* | 0.44 | ~68% | 0.37 | <68% | 0.49 | 68-95% |
| Cycle 3 | 2.3E-05 | 483 | 1051 | *31.5* | 0.93 | 95-99% | 0.97 | 95-99% | 0.84 | 95-99% |
| Cycle 4 | 0.22 | 478 | 1041 | *31.5* | 0.27 | <68% | -0.06 | <68% | 0.40 | ~68% |
| C2+C3+C4 | 1.8E-05 | 1373 | 3007 | *31.4* | 0.49 | 95-99% | 0.45 | 95-99% | 0.56 | >99% |
| | By Cycle & Summed, Chile | | | | | | | | | |
| Cycle 2 | 0.080 | 24 | 63 | *27.6* | 1.35 | <68% | 1.38 | <68% | 1.23 | ~68% |
| Cycle 3 | 0.13 | 15 | 94 | *13.8* | 1.12 | <68% | 1.07 | ~68% | 0.44 | <68% |
| Cycle 4 | 0.48 | 19 | 76 | *20.0* | 1.01 | <68% | 0.75 | <68% | 0.07 | <68% |
| C2+C3+C4 | 0.0068 | 58 | 233 | *19.9* | 1.03 | 68-95% | 1.06 | 68-95% | 0.36 | 68-95% |
| | By Cycle & Summed, East Asia | | | | | | | | | |
| Cycle 2 | 0.27 | 64 | 200 | *24.2* | 1.79 | 68-95% | -0.20 | <68% | 0.14 | <68% |
| Cycle 3 | 0.32 | 79 | 204 | *27.9* | 0.62 | <68% | 0.40 | <68% | 0.05 | <68% |
| Cycle 4 | 0.48 | 84 | 248 | *25.3* | 0.22 | <68% | 1.12 | <68% | 0.06 | <68% |
| C2+C3+C4 | 0.13 | 227 | 652 | *25.8* | 0.76 | 68-95% | 0.34 | <68% | 0.19 | <68% |
| | By Cycle & Summed, Europe | | | | | | | | | |
| Cycle 2 | 0.87 | 187 | 351 | *34.8* | -0.12 | <68% | -0.14 | <68% | 0.34 | <68% |
| Cycle 3 | 0.036 | 230 | 411 | *34.9* | 0.81 | 68-95% | 1.19 | 68-95% | 0.27 | <68% |
| Cycle 4 | 0.090 | 240 | 396 | *37.7* | 0.75 | 68-95% | 0.36 | <68% | 0.38 | <68% |
| C2+C3+C4 | 0.023 | 657 | 1158 | *36.2* | 0.44 | 68-95% | 0.63 | 68-95% | 0.36 | 68-95% |
| | By Cycle & Summed, North America | | | | | | | | | |
| Cycle 2 | 0.016 | 137 | 301 | *31.3* | 1.09 | 68-95% | 0.70 | ~68% | 0.96 | 68-95% |
| Cycle 3 | 0.0014 | 159 | 342 | *31.7* | 0.99 | ~95% | 1.01 | ~95% | 1.10 | 68-95% |
| Cycle 4 | 0.62 | 136 | 320 | *29.6* | 0.12 | <68% | -0.76 | <68% | 0.38 | <68% |
| C2+C3+C4 | 0.0015 | 431 | 964 | *30.9* | 0.64 | 95-99% | 0.45 | 68-95% | 0.86 | ~95% |

Table 1 Notes: [1]>99% AD-test confidence: dark purple; 95-99%: mid purple; 90-95%: light purple; 80-90%: faintest purple. [2] F-M: quartile rank for Female PIs minus rank for male PIs. [3] CL: Confidence Level for rank difference. Male or Female dominance >0.2 in rank difference and/or >68% confidence level highlighted blue (male) or tan (female) with deepening shading for larger |F-M| values (0.2-0.5; 0.5-1.0; >1.0) and for more significant CL values (68-95%; 95-99%; >99%).

**TABLE 2: ALMA Cycles 0 & 1: AD Tests for Rank Distributions and Ranks at the Three Quartiles**

| ALMA | Anderson-Darling Test p-value[1] | NUMBER OF PIS | | % F PIs | 25th Percentile | | 50% Percentile | | 75th Percentile | |
|---|---|---|---|---|---|---|---|---|---|---|
| | | F | M | | F-M[2] | CL[3] | F-M | CL | F-M | CL |
| | Top 18% for Cycle 0, all regions combined | | | | | | | | | |
| Cycle 0 | 0.19 | 41 | 122 | *25.2* | -1.05 | <68% | 0.28 | <68% | -0.86 | ~68% |
| | Top 70% for Cycle 1, all regions combined | | | | | | | | | |
| Cycle 1 | 0.18 | 209 | 567 | *26.9* | 0.10 | <68% | 0.48 | <68% | -0.26 | <68% |

Table 2 Notes: Columns and shading as for Table 1

## TABLE 3: NRAO: AD Tests for Rank Distributions and Ranks at the Three Quartiles

| NRAO | Anderson-Darling Test p-value[1] | NUMBER OF PIs | | % F PIs | 25th percentile | | 50th percentile | | 75th percentile | |
|---|---|---|---|---|---|---|---|---|---|---|
| | | F | M | | F-M[2] | CL[3] | F-M | CL | F-M | CL |
| | **By Year & Summed** | | | | | | | | | |
| **2012A+B** | 0.41 | 192 | 528 | *26.7* | 0.23 | <68% | 0.39 | <68% | 0.05 | <68% |
| **2013A+B** | 0.18 | 194 | 570 | *25.4* | -0.03 | <68% | -0.53 | <68% | -0.78 | 68-95% |
| **2014A+B** | 0.047 | 242 | 558 | *30.3* | 0.60 | 68-95% | 0.40 | <68% | -0.14 | <68% |
| **2015A+B** | 0.67 | 168 | 481 | *25.9* | -0.07 | <68% | -0.15 | <68% | 0.003 | <68% |
| **2016A+B** | 0.13 | 164 | 455 | *26.5* | 0.86 | 68-95% | 0.38 | <68% | -0.03 | <68% |
| **2012A-2016B** | 0.017 | 960 | 2592 | *27.0* | 0.28 | 68-95% | 0.11 | <68% | -0.17 | <68% |
| | **By Semester** | | | | | | | | | |
| **2012A** | 0.91 | 108 | 291 | *27.1* | 0.19 | <68% | 0.14 | <68% | -0.17 | <68% |
| **2012B** | 0.40 | 84 | 237 | *26.2* | 0.71 | <68% | 0.88 | <68% | 0.26 | <68% |
| **2013A** | 0.31 | 107 | 306 | *25.9* | -0.13 | <68% | -0.43 | <68% | -0.81 | 68-95% |
| **2013B** | 0.51 | 87 | 264 | *24.8* | -0.04 | <68% | -0.68 | <68% | -0.64 | <68% |
| **2014A** | 0.57 | 134 | 310 | *30.2* | 0.01 | <68% | 0.21 | <68% | -0.17 | <68% |
| **2014B** | 0.049 | 108 | 248 | *30.3* | 1.02 | 68-95% | 0.56 | <68% | -0.04 | <68% |
| **2015A** | 0.73 | 93 | 282 | *24.8* | -0.27 | <68% | -0.45 | <68% | -0.09 | <68% |
| **2015B** | 0.49 | 75 | 199 | *27.4* | 1.07 | <68% | 0.38 | <68% | 0.77 | <68% |
| **2016A** | 0.65 | 96 | 222 | *30.2* | 0.47 | <68% | -0.27 | <68% | -0.18 | <68% |
| **2016B** | 0.017 | 68 | 233 | *22.6* | 1.48 | ~95% | 1.42 | 68-95% | 0.17 | <68% |
| **2017A** | 0.13 | 101 | 209 | *32.6* | 0.55 | ~68% | 1.06 | 68-95% | 0.57 | <68% |
| | **By Telescope** | | | | | | | | | |
| **VLA** | 0.095 | 666 | 1621 | 29.1 | 0.34 | 68-95% | 0.08 | <68% | -0.23 | <68% |
| **VLBA** | 0.87 | 79 | 408 | 16.2 | 0.01 | <68% | 0.51 | <68% | 0.27 | <68% |
| **GBT** | 0.11 | 215 | 563 | 27.6 | 0.39 | <68% | 0.00 | <68% | -0.32 | <68% |

Table 3 Notes: Columns and shading as for Table 1

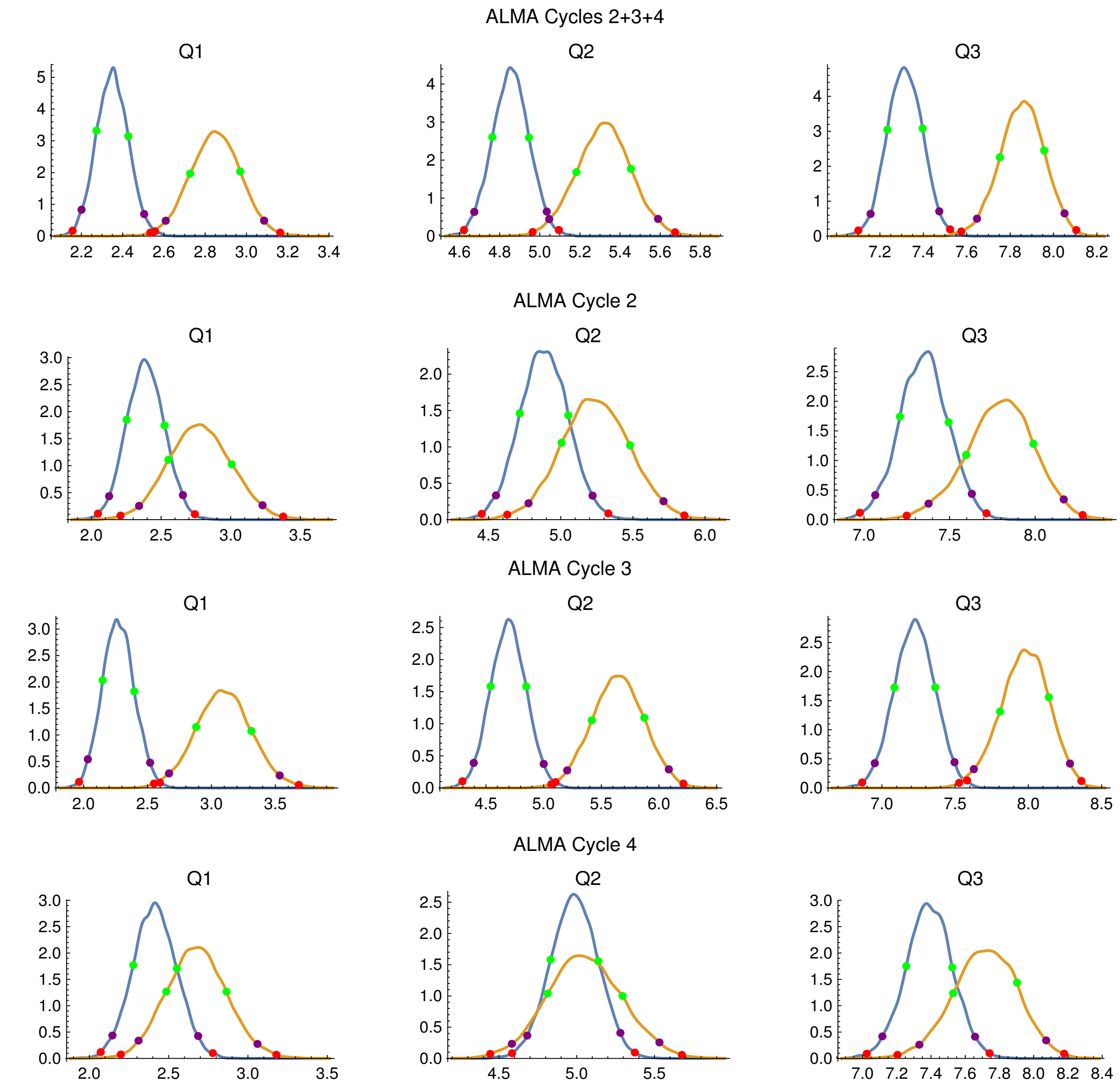

**Figure 1**: Modeled distributions of the 25th, 50th and 75th percentile values for normalized rank (1-10), derived via bootstrapping, for ALMA Cycle 2, 3 & 4 combined and each of the three cycles separately. Blue curves: male PIs; red curves: female PIs. Green, purple and red dots delimit, respectively, the 68%, 95% and 99% probability intervals. For the combined results (top panel) the 25th, 50th and 75th percentile values for men are found to be better than for women at the ~99%, >95% and >99% confidence level respectively. The most significant effect occurs in Cycle 3.

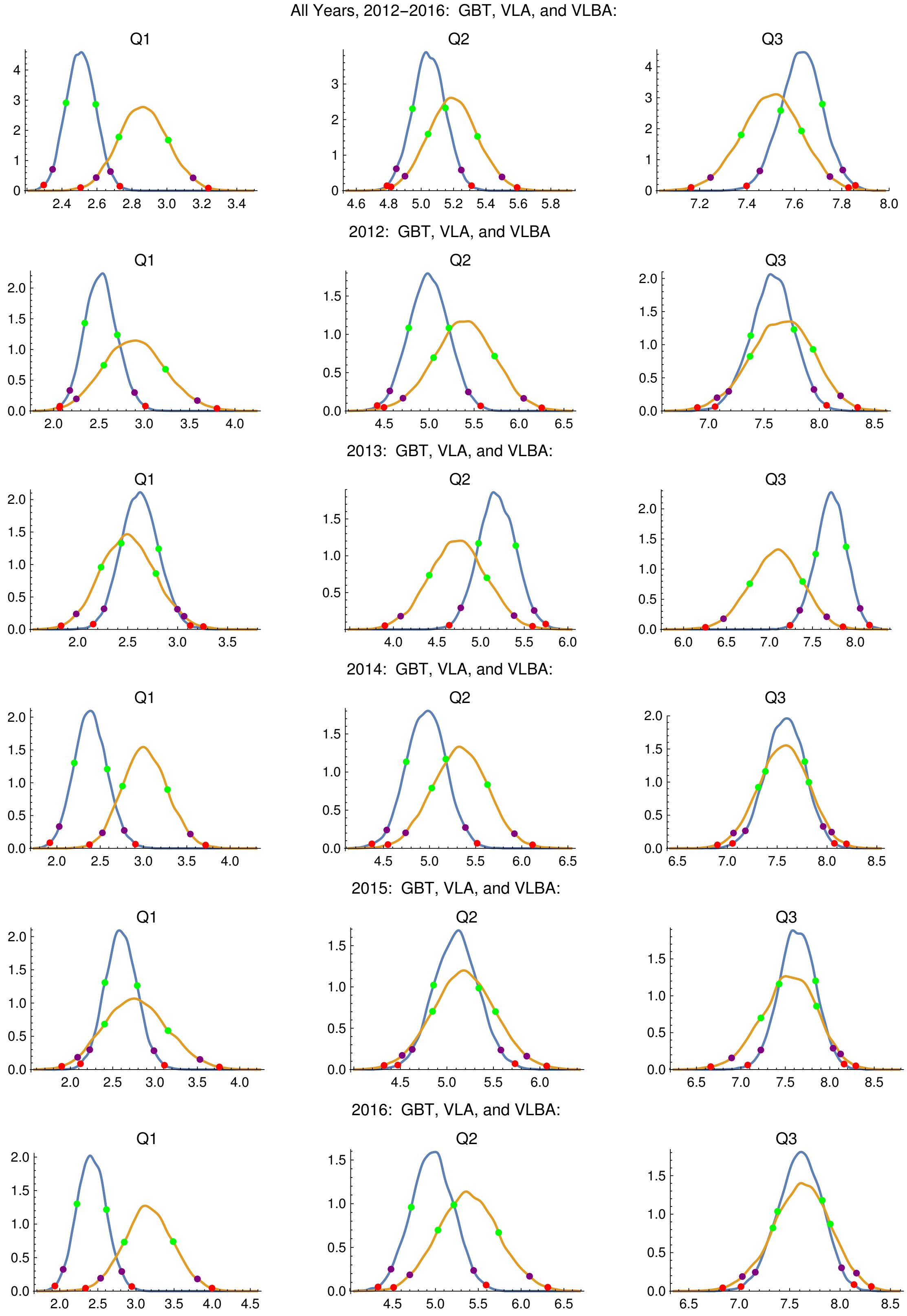

**Figure 2:** As for Figure 1 but for NRAO years 2012-2016 combined and each year separately. A trend for male dominance may be present overall, but with much lower significance than for ALMA.

**TABLE 4: SRP Panel Member Continuations Between Consecutive NRAO Semesters**

| Members in common | 12A-12B | 12B-13A | 13A-13B | 13B-14A | 14A-14B | 14B-15A | 15A-15B | 15B-16A | 16A-16B |
|---|---|---|---|---|---|---|---|---|---|
| Men | 22 | 33 | 15 | 28 | 22 | 30 | 33 | 49 | 21 |
| Women | 6 | 10 | 7 | 13 | 8 | 7 | 4 | 11 | 5 |
| Total | 28 | 42 | 22 | 41 | 30 | 37 | 37 | 60 | 26 |

## V. Discussion

### a. Gender-Related Trends

For ALMA Cycles 2-4, the results for quartile 1 are of most importance in terms of how gender-related systematics will reflect on proposal success, since only 25% or less of ALMA proposals get observed. For NRAO both quartiles 1 and 2 are of importance since the over-subscription rate is lower than for ALMA. It is relevant to investigate the trends across the full range of the rankings as gender-related trends can show up across the whole range, however it should be kept in mind that female dominance in Q3 is not a good indicator of trend reversal since it may trace proposals that have been shifted down to below the 50$^{th}$ percentile by male dominance in the top 50%.

For ALMA the trends seen in the combined data are strong, and it is hard to avoid the conclusion that they are real. They all occur in the same sense – men outperform women – and for all 3 quartiles, as evidenced by the blue shading (male-dominated quartiles) in Table 1 and small number of tan shaded cells (female-dominated quartiles). The overall confidence levels can reach over 99.99% and the quartile difference reliabilities over 95%. When broken down by cycle it is clear that the trend is strongest in Cycle 3.

When the ALMA data are broken by region, the same general trends are evident overall. There is least evidence for the trends favoring male PIs for East Asia, while the strongest signal comes from Chile and North America. The result that the trend is most significant for Cycle 3 is driven by Europe and North America.

For NRAO there is lesser significance for male PIs outperforming female PIs, however the results do point to an overall trend favoring men. This is best appreciated from the patterns of blue (male-dominated quartiles) vs. tan (female-dominated quartiles) in Table 3. There is evidence that the trend sometimes evaporates, with 3 semesters showing some evidence favoring women instead. The AD test results confirm a difference in the rank distributions for all data combined with confidence level 98%, and there are two individual semesters with significance above 95%: 2014B and 2016B.

The breakdown of the results for NRAO by telescope provides limited insights, given that the number statistics are not high. A gender-related signal may be present for VLA and for GBT. The data for VLBA are too sparse to allow any conclusions.

Our assessment of the SRP gender fractions for NRAO shows no evidence that this is related to gender-related trends in proposal outcomes. Nor is there evidence that the changing gender-related proposal outcome trend from 2014A to 2014B and from 2016A to 2016B are associated with a high turnover rate in SRP membership.

A difference is seen in the most recent trends for ALMA and NRAO, namely that in Cycle 4 the ALMA male-dominance trend appears to become *less* significant, while for NRAO semesters 2016B and 2017A the trend appears to become *more* significant. In both cases the review panel members had been alerted to the discovery of gender-related effects in prior cycles.

### b. The Influence of Seniority on Gender-Related Trends

The recently published analyses of gender-related trends in the proposal review outcomes of two large observatories, HST (Reid 2014) and ESO (Patat 2016), have inferred that the seniority of the PI is a significant factor affecting the proposal review results. Patat in particular concludes that the strongest result of the ESO analysis is that it is the more senior scientists who are more likely to be successful in the review process. He then discusses the possibility that the apparent gender-related trends he observes could be largely attributable to the lower F/M gender fraction amongst senior PIs in the proposal pool compared to early career PIs. As was mentioned in Section I of this paper, a bias of this nature is referred to as prestige bias in the Social Sciences community: the more well established members of a community are perceived to be capable of superior work. To some degree prestige bias is built into many telescope review processes because review panels are often advised to take into account their perception of the ability of the proposing team to undertake the proposed study. If the more prestigious group is male-dominated due to external factors, a gender bias will result.

On the other hand, Reid (2014) has noted that there is some evidence in the HST proposal review data that it may be the more junior female astronomers that have a better chance of a good proposal outcome, and that it is the proposal outcomes of the more senior female PIs that drive the apparent gender-related trends towards poorer outcomes for women. Such a result is contrary to the expected effect of prestige bias, if we assume that telescope review panels accord senior astronomers more consideration *regardless of their gender*. If the HST result bears out under further scrutiny then the situation is clearly more complex than discussed by Patat; in particular there is evidence that the younger generation of astronomers is more gender-neutral than older generations, which may imply that the more senior female PIs are still be being held to higher standards than their male contemporaries.

It is unfortunate that at the present time we do not have data on the seniority of the ALMA and NRAO user communities, therefore the current analysis cannot shed further light on this important question at this time.

## VI. Summary

It has been shown that the ALMA proposal review process in Cycles 2-4 has resulted in a significant gender-related systematic trend favoring men vs. women, with a reliability of 99.998% that the ranking distributions are not the same. The result is consistent in all cycles however it is most significant in Cycle 3, which dominates the combined Cycle 2-4 data. The result is also consistent across all four regions however it is least significant for East Asia.

For the NRAO telescopes we also find that an overall signal, favoring men, significant at the 98% confidence level. It is strongest in semesters 2014B and 2016B with significance 95% and 98% respectively, however it is weaker in semesters 2013A, 2013B and 2015A. For the 3 individual NRAO telescopes there is a ~90% confidence signal that male PIs do better than female PIs on both GBT and VLA, when the data from all semesters are combined. The data for VLBA are too sparse for reliable conclusions. We find no evidence that the gender balance on the NRAO SRPs has any relation to the observed systematics.

The NRAO is concerned by the evidence of gender-related bias revealed by this study, and is committed to improve the level of equity and inclusion in its review processes. The observatory will continue to investigate the effects and the root causes of gender-related systematics, and will explore mechanisms to ensure greater equity and transparency, including providing information to review panel members and the astronomy community regarding these issues.

Acknowledgements
ALMA is a partnership of ESO (representing its member states), NSF (USA) and NINS (Japan), together with NRC (Canada), NSC and ASIAA (Taiwan), and KASI (Republic of Korea), in cooperation with the Republic of Chile. The Joint ALMA Observatory is operated by ESO, AUI/NRAO and NAOJ. The National Radio Astronomy Observatory is a facility of the National Science Foundation operated under cooperative agreement by Associated Universities, Inc.

## References

Babu, G. J. & Feigelson, E.D. 2006 Astrostatistics: Goodness-of-fit and all that!, in *Astronomical Data Analysis Software and Systems XV* (eds. C. Gabriel et al.), ASP Conf. #351, 127

Efron, B., 1982, The Jackknife, the Bootstrap and Other Resampling Plans, CBMS–NSF Regional Conference Series in Applied Mathematics (Book 38), Society for Industrial and Applied Mathematics, Philadelphia, PA

Patat, F. 2016, Gender Systematics in Telescope Time Allocation at ESO, arXiv:1610.00920v1 [physics.sco-ph]

Reid, I. N. 2014, Gender-based Systematics in HST Proposal Selection, PASP, 126, 923

# Appendix

We show more detailed graphical results for the combined Cycle 2-4 data for ALMA and the 2012-2016 data for all telescopes combined for NRAO. First we report the number of PIs in the sample, and then 25th, 50th and 75th percentile for the male and the female PIs (replicated in Tables 1 and 3 in the main article), with their upper and lower 1$\sigma$ uncertainties. We then show the full histograms of the normalized and linearized ranks and give the Anderson-Darling two sample p-values for a comparison of these histograms. The density estimator fit to these distributions is shown as a solid line, magenta for females and blue for males. We then, for each gender, plot the distribution of quartile values obtained by resampling (generating 10,000 replicates of the original data set) from the smooth approximating probability density function. The green, purple, and red dots delimit, respectively, the 68%, 95%, and 99% probability intervals.

The full set of statistical results is available at:
https://science.nrao.edu/science/reports/StatisticalData

# Figure A1: Results for ALMA Cycles 2, 3 and 4 Combined

| 7

ALMA Cycle 2+3+4 Data

No. Female PIs = 1373

No. Male PIs = 3007

F PI Quartiles: {2.83884, 5.3092, 7.85877}

M PI Quartiles: {2.34598, 4.85671, 7.29988}

Anderson-Darling two-sample test p-value: 0.0000182212

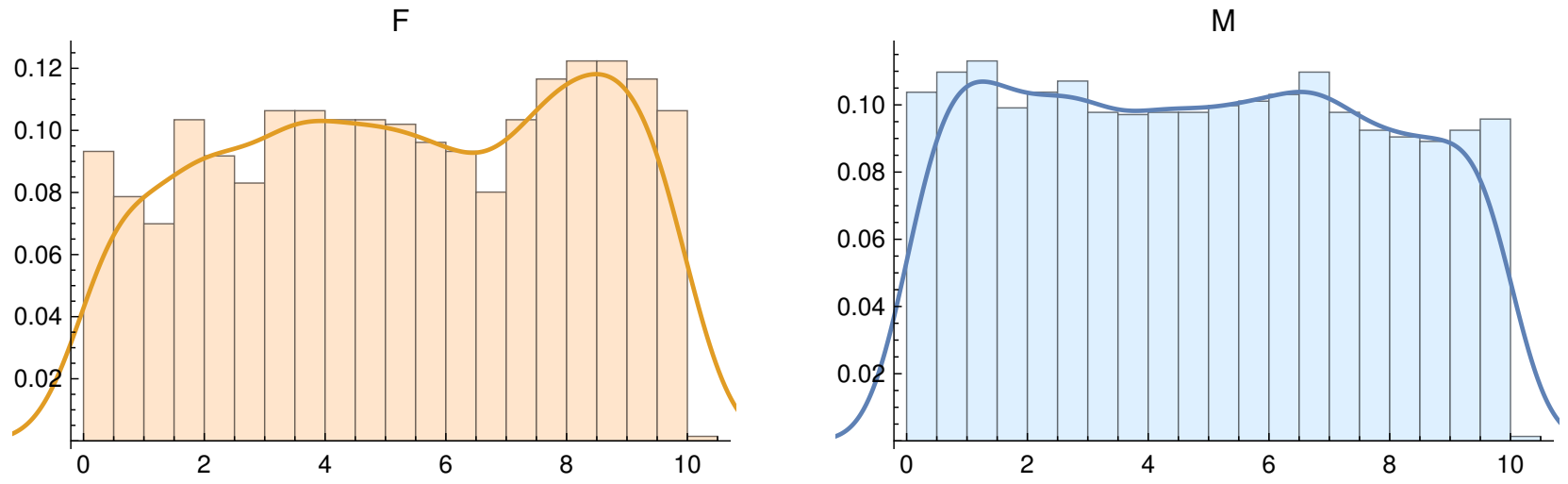

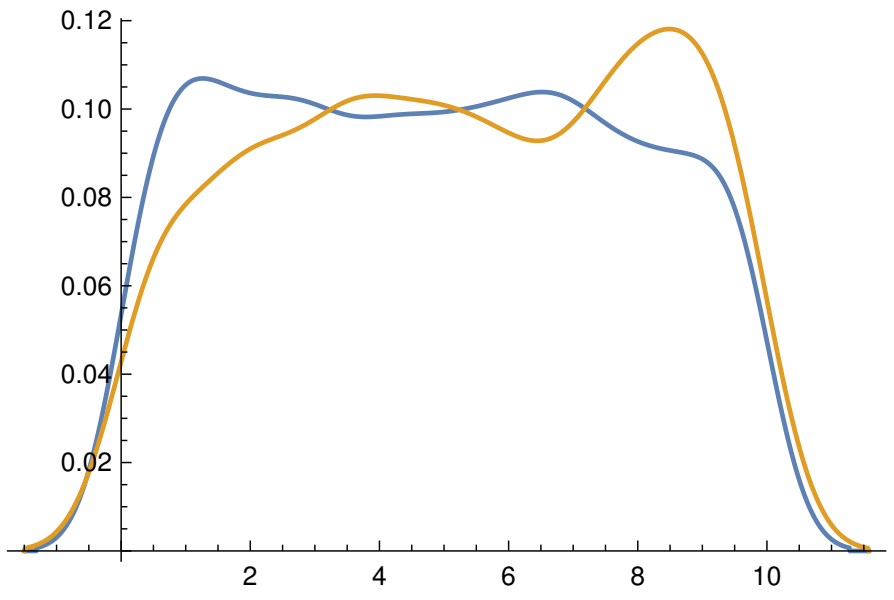

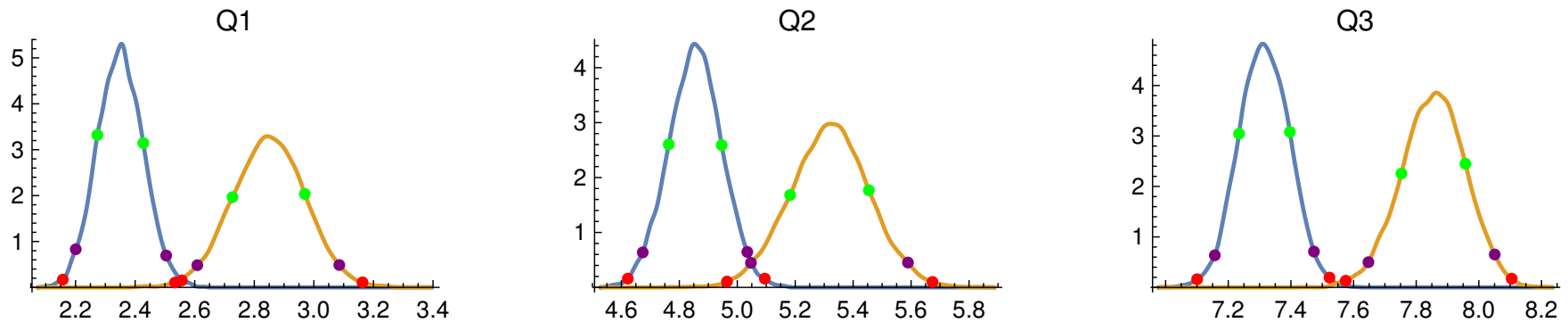

# Figure A2: Results for NRAO, Years 2012-2016, all Telescopes Combined

All Years, 2012-2016: GBT, VLA, and VLBA

No. Female PIs = 960

No. Male PIs = 2592

F PI Quartiles: {2.775, 5.195, 7.46}

M PI Quartiles: {2.5, 5.085, 7.625}

Anderson-Darling two-sample test p-value: 0.0167233

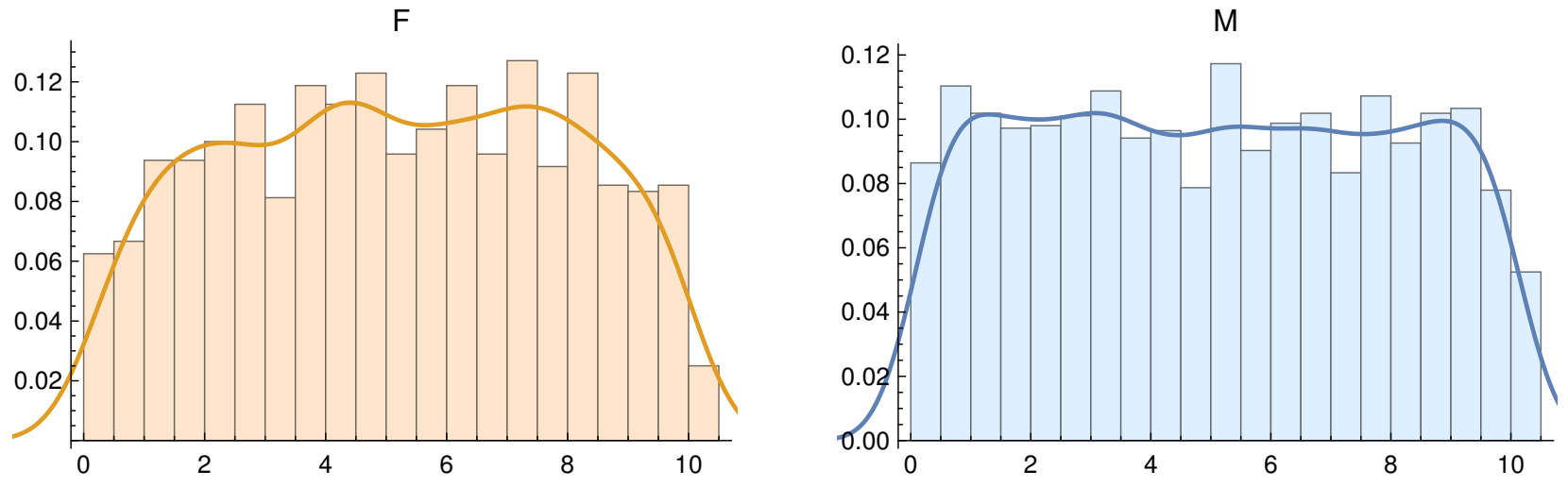

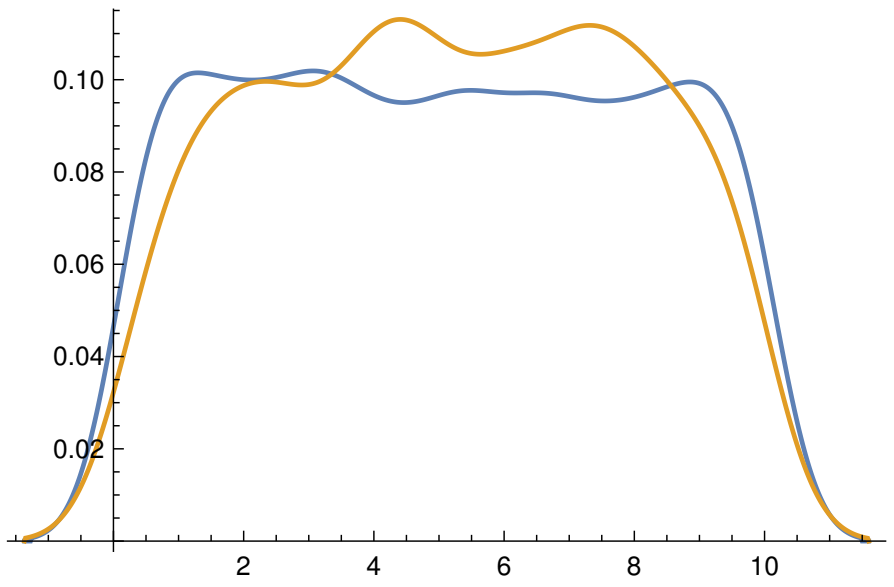

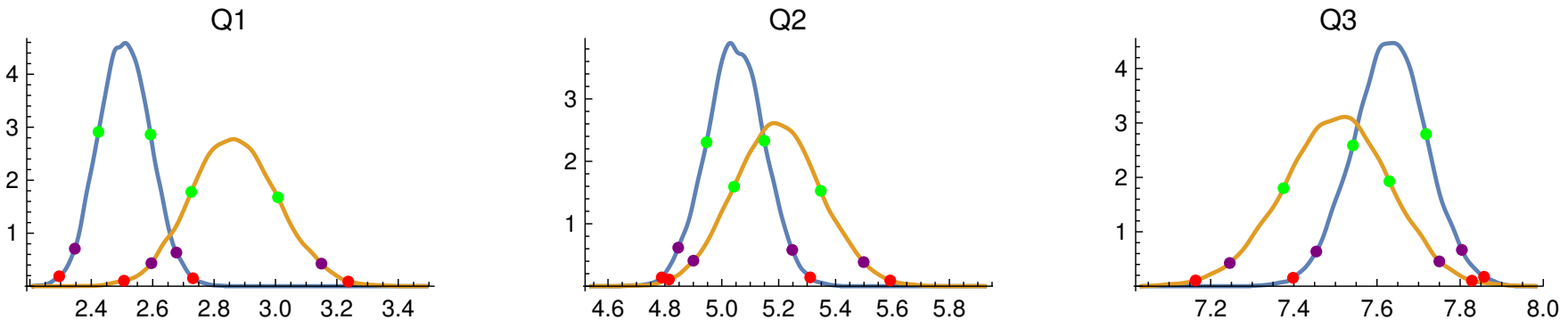